\documentclass[preprint,prd,aps,nofootinbib]{revtex4}
\usepackage{graphicx}
\usepackage{color}

\begin{document}

%\preprint{hep-ph/0412147}

\title{A mechanism relating the fermionic mass hierarchy to the flavor mixing}

\author{Jin-Lei Yang$^{1,2,3}$\footnote{jlyang@hbu.edu.cn},Hai-Bin Zhang$^{1,2,3}$\footnote{hbzhang@hbu.edu.cn},Tai-Fu Feng$^{1,2,3}$\footnote{fengtf@hbu.edu.cn}}

\affiliation{Department of Physics, Hebei University, Baoding, 071002, China$^1$\\
Key Laboratory of High-precision Computation and Application of Quantum Field Theory of Hebei Province, Baoding, 071002, China$^2$\\
Research Center for Computational Physics of Hebei Province, Baoding, 071002, China$^3$}

\begin{abstract}
Considering the hierarchical structure of fermionic masses and the fermionic flavor mixing puzzles in the Standard Model, we propose to relate them by the see-saw mechanism, i.e. only the third generation of quarks and charged leptons achieve the masses at the tree level, the first two generations achieves masses through the mixings with the third generation, and the neutrinos achieve tiny Majorana masses by the so-called Type-I see-saw mechanism. This new picture at the fermion sector can explain simultaneously the flavor mixing puzzle and mass hierarchy puzzle in the SM. In addition, a flavor-dependent model (FDM) is proposed to realize the new mechanism, and observing the top quark rare decay processes $t\to ch$, $t\to uh$ and the lepton flavor violation processes $\mu\to3e,\;\tau\to3e,\;\mu\to3\mu$ is effective to test the proposed FDM.
\end{abstract}

\maketitle

The Standard Model (SM) achieves great success in describing the weak, strong, and electromagnetic gauge interactions of fundamental particles. However, the Yukawa couplings in the SM are still enigmatic so far because the large hierarchical structure of masses across the three families
\begin{eqnarray}
&&\frac{m_t}{m_c}\approx104,\;\frac{m_c}{m_u}\approx773,\;\frac{m_b}{m_s}\approx51,\;\frac{m_s}{m_d}\approx20,\;\frac{m_\tau}{m_\mu}\approx17,\;\frac{m_\mu}{m_e}\approx207,
\end{eqnarray}
while they share common SM gauge group quantum numbers. In addition, the mixing pattern of quarks described by the Cabibbo-Kobayashi-Maskawa (CKM) matrix~\cite{Cabibbo:1963yz,Kobayashi:1973fv} is not predicted from first principles in the SM, and the nonzero neutrino masses and mixings observed at the neutrino oscillation experiments~\cite{ParticleDataGroup:2022pth} make the so-called flavor puzzle in the SM more acutely. It indicates the existing of a more fundamental theory beyond the SM to well explain the large hierarchical structure of fermionic masses, flavor mixings and nonzero neutrino masses. There are some attempts to explain the fermionic mass hierarchies and flavor mixings. For example, the authors of Refs.~\cite{Froggatt:1978nt,Koide:1982ax,Leurer:1992wg,Ibanez:1994ig,Babu:1995hr} try to explain the fermionic flavor mixings by imposing the family symmetries, some proposals to explain the fermionic mass hierarchies can be found in Refs.~\cite{Randall:1999ee,Kaplan:2001ga,Chen:2008tc,Buras:2011ph,King:2013eh,King:2014nza,King:2015aea,King:2017guk,Weinberg:2020zba,Feruglio:2019ybq}. These studies illustrate that explaining the observed fermionic mass spectrum and mixings is one of the most enigmatic questions in particle physics, which may help to understand the flavor nature and seek possible new physics (NP).

Weinberg made important efforts to the fermionic mass spectrum~\cite{Weinberg:1971nd}, flavor problem in the SM and the nonzero neutrino masses. He was the first to propose the Type-I see-saw mechanism to give the tiny neutrino masses naturally~\cite{Weinberg:1979sa} which is one of the most popular mechanisms so far to arise the Majorana neutrino masses. In 2020~\cite{Weinberg:2020zba}, Weinberg proposed a class of models in which only the third generation of quarks and leptons achieve the masses at the tree level, while the masses for the second and first generations are produced by one-loop and two-loop radiative corrections respectively, although he pointed that these models are not realistic for some reasons. Adopting Weinberg's ideas of see-saw mechanism and only the third generation of quarks, leptons achieve the masses at the tree level, we propose to apply the see-saw mechanism to the mass matrices of quarks and leptons, in which the second and first generations achieve masses through the mixings with the third generation, i.e. the mass matrices of quarks and leptons can be written as
\begin{eqnarray}
&&m_q=\left(\begin{array}{ccc} 0 & m_{q,12} & m_{q,13}\\
m_{q,12}^* & 0 & m_{q,23}\\
m_{q,13}^* & m_{q,23}^* & m_{q,33}\end{array}\right),m_e=\left(\begin{array}{ccc} 0 & m_{e,12} & m_{e,13}\\
m_{e,12}^* & 0 & m_{e,23}\\
m_{e,13}^* & m_{e,23}^* & m_{e,33}\end{array}\right),m_\nu=\left(\begin{array}{cc} 0 & M_D^T\\
M_D & M_R\end{array}\right),\label{eq2}
\end{eqnarray}
where $q=u,d$, $m_{q,33}$ and $m_{e,33}$ are real, $M_D,\;M_R$ are the $3\times3$ Dirac and Majorana mass matrices respectively (the nonzero neutrino masses are obtained by the so-called Type I see-saw mechanism). Then considering the measured fermionic masses, under the approximation $|m_{f,12}|,\;|m_{f,13}|,\;|m_{f,23}|\ll |m_{f,33}|\;(f=u,\;d,\;e)$ one can obtain
\begin{eqnarray}
&&m_{f,33}=m_{f_3}-m_{f_1}-m_{f_2},\nonumber\\
&&|m_{f,23}|=[(m_{f_1}+m_{f_2})m_{f,33}-|m_{f,13}|^2]^{1/2},\nonumber\\
&&|m_{f,12}|=\frac{|m_{f,13}||m_{f,23}|}{|m_{f,33}|} \cos(\theta_{f,12}+\theta_{f,23}-\theta_{f,13})\nonumber\\
&&\qquad\qquad+\Big\{[\frac{|m_{f,13}||m_{f,23}|}{|m_{f,33}|^2}\cos(\theta_{f,12}+\theta_{f,23}-\theta_{f,13})]^2+\frac{m_{f_1}m_{f_2}}{|m_{f,33}|^2}\Big\}^{1/2}|m_{f,33}|,\label{eq3}
\end{eqnarray}
where $\theta_{f,ij}\;(ij=12,\;13,\;23)$ are defined as $m_{f,ij}=|m_{f,ij}|e^{i\theta_{f,ij}}$, and $m_{f_k}\;(k=1,\;2,\;3)$ is $k-$generation fermion $f$ mass. Eq.~(\ref{eq2}) shows that the fermionic flavor mixings are related to the fermionic mass hierarchies after applying the see-saw mechanism to the fermionic mass matrices. In this case, the measured quark masses, lepton masses and their flavor mixings (CKM matrix at the quark sector and PMNS matrix at the lepton sector) will impose strict constraints on the parameters in Eq.~(\ref{eq2}). Due to the neutrino masses are not fixed experimentally, we focus on the flavor mixings at the quark sector firstly\footnote{The analysis about the mass structure and flavor mixings at the lepton sector are similar to the one at the quark sector, and the free parameters in the matrices $m_e,\;m_\nu$ defined in Eq.~(\ref{eq2}) are constrained less strictly than the quark sector because the nonzero neutrino masses are obtained by the Type I see-saw mechanism.}.

As mentioned above, the mixing pattern of quarks is described by the CKM matrix, and quarks' flavor mixing is related to the quarks' mass hierarchies by the new definitions of fermionic mass matrices in Eq.~(\ref{eq2}). Taking the measured quark masses as input, the remaining free parameters in the matrix $m_q$ are $|m_{q,13}|,\;\theta_{q,ij}\;(q=u,\;d)$, and they are required to fit the measured CKM matrix~\cite{ParticleDataGroup:2022pth}
\begin{eqnarray}
&&|V_{\rm CKM}|=\left(\begin{array}{ccc} 0.97435\pm0.00016 & 0.22500\pm0.00067 & 0.00369\pm0.00011\\
0.22486\pm0.00067 & 0.97349\pm0.00016 & 0.04182^{+0.00085}_{-0.00074}\\
0.00857^{+0.00020}_{-0.00018} & 0.04110^{+0.00083}_{-0.00072} & 0.999118^{+0.000031}_{-0.000036}\end{array}\right).
\end{eqnarray}

\begin{figure}
\setlength{\unitlength}{1mm}
\centering
\includegraphics[width=2.1in]{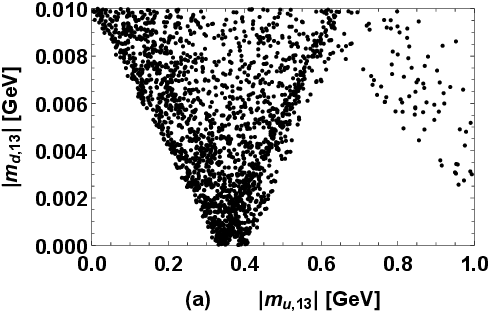}
\vspace{0.5cm}
\includegraphics[width=2.1in]{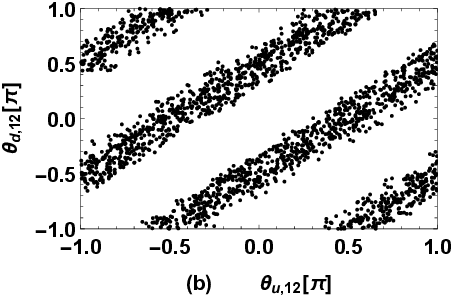}
\vspace{0.5cm}
\includegraphics[width=2.1in]{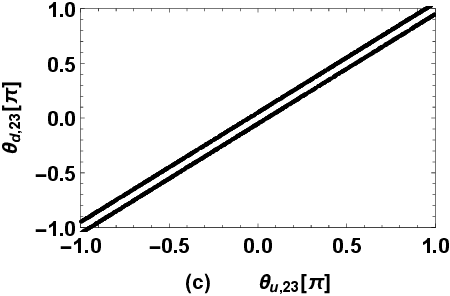}
\vspace{-0.4cm}
\caption[]{Scanning the parameter space in Eq.~(\ref{eq5}) and keeping the CKM matrix elements in the ranges shown in Eq.~(\ref{eq6}), the allowed ranges of $|m_{d,13}|-|m_{u,13}|$ (a), $\theta_{d,12}-\theta_{u,12}$ (b), $\theta_{d,23}-\theta_{u,23}$ (c) are plotted.}
\label{Fig1}
\end{figure}
In order to see whether the mass matrices defined in Eq.~(\ref{eq2}) can fit the measured CKM matrix, we take the pole quark masses $m_t=174.3\;{\rm GeV},\;m_c=1.67\;{\rm GeV},\;m_u=2.16\;{\rm MeV},\;m_b=4.78\;{\rm GeV},\;m_s=93.4\;{\rm MeV},\;m_d=4.67\;{\rm MeV}$ as inputs and scan the free parameters $|m_{q,13}|,\;\theta_{q,ij},\;(ij=12,\;13,\;23)$ in the following range
\begin{eqnarray}
&&|m_{u,13}|=(0.0\sim1.0)\;{\rm GeV},\;|m_{d,13}|=(0.0\sim0.1)\;{\rm GeV},\;\theta_{q,ij}=(-\pi\sim\pi),\label{eq5}
\end{eqnarray}
and keep
\begin{eqnarray}
&&|V_{\rm CKM,12}|=(0.224\sim0.226),\;|V_{\rm CKM,21}|=(0.22386\sim0.22586),\nonumber\\
&&|V_{\rm CKM,23}|=(0.04082\sim0.04282),\;|V_{\rm CKM,32}|=(0.0401\sim0.0421),\nonumber\\
&&|V_{\rm CKM,13}|=(0.00359\sim0.00379),\;|V_{\rm CKM,31}|=(0.00847\sim0.00867),\label{eq6}
\end{eqnarray}
in the scanning. The allowed ranges of $|m_{d,13}|-|m_{u,13}|$, $\theta_{d,12}-\theta_{u,12}$, $\theta_{d,23}-\theta_{u,23}$ are plotted in Fig.~\ref{Fig1} (a), (b), (c) respectively. The picture shows that the mass matrices proposed in this work can well coincide with the measured quark masses and CKM matrix. And to fit the measured CKM matrix elements, the values of $|m_{d,13}|$, $\theta_{d,23}$ are related to the chosen values of $|m_{u,13}|$, $\theta_{u,23}$ respectively, and $|\theta_{d,23}|-|\theta_{u,23}|\approx\pm(0.04\sim0.06)\pi$ as shown in Fig.~\ref{Fig1} (c).

The analysis above illustrate that applying the see-saw mechanism to the quarks' mass matrices and the first two generations achieve masses through the mixings with the third generation can well fit the measured quark masses and CKM matrix. The fermionic flavor mixings are related to the fermionic mass hierarchies in this new mechanism which can explain simultaneously the flavor mixings puzzle and mass hierarchy puzzle in the SM. This mechanism can be realized simply by setting the Yukawa couplings $Y_{q,11}=Y_{q,22}=0$ in the SM, but it is trivial to set the terms allowed by the gauge symmetries to zero. On the other hand, the mass matrices defined in Eq.~(\ref{eq2}) can be realized in a NP model by extending the SM with an extra $U(1)_F$ local gauge group where the corresponding $U(1)_F$ charges are related to the particles' flavor, i.e. the gauge group of such flavor-dependent $U(1)_F$ model (FDM) is $SU(3)_C\otimes SU(2)_L\otimes U(1)_Y\otimes U(1)_F$. The gauge charges of fermions corresponding to $SU(2)_L\otimes U(1)_Y\otimes U(1)_F$ read
\begin{eqnarray}
&&L_{1}\sim(2,Y_{L},z),L_{2}\sim(2,Y_{L},-z),L_{3}\sim(2,Y_{L},0),\nonumber\\
&&R_{1}\sim(1,Y_{R},-z),R_{2}\sim(1,Y_{R},z),R_{3}\sim(1,Y_{R},0),\nonumber\\
&&\nu_{R_1}\sim(1,0,-z),\nu_{R_2}\sim(1,0,z),\label{eq7}
\end{eqnarray}
where the nonzero $z$ denotes the $U(1)_F$ charge, $L_i$ ($L=l,\;q$) denote the $i-$generation left-handed fermion doublets in the SM, $R_{i}$ ($R=u_R,\;d_R,\;e_R$) denotes the $i-$generation right-handed fermion singlets in the SM, $\nu_{R_i}\;(i=1,2)$ is the right-handed neutrinos\footnote{The third generation of right-handed neutrino $\nu_{R_3}$ has zero $U(1)_F,\;U(1)_Y$ charges which is trivial, hence only two generations of right-handed neutrinos are introduced.}, $Y_L,\;Y_R$ are the $U(1)_Y$ charges corresponding to doublets $L$ and singlets $R$ in the SM respectively. The chiral anomaly cancellation can be guaranteed for the fermionic charges defined in Eq.~(\ref{eq7}). The scalar sector is extended by introducing two new doublets and one new singlet
\begin{eqnarray}
&&\Phi_{1}\sim(2,\frac{1}{2},z),\Phi_{2}\sim(2,\frac{1}{2},-z),\Phi_{3}\sim(2,\frac{1}{2},0),\chi\sim(1,0,2z),
\end{eqnarray}
with $\Phi_{3}$ corresponding to the SM Higgs doublet. The scalar potential in the FDM reads
\begin{eqnarray}
&&V=-M_{\Phi_1}^2 \Phi_1^\dagger\Phi_1-M_{\Phi_2}^2 \Phi_2^\dagger\Phi_2-M_{\Phi_3}^2 \Phi_3^\dagger\Phi_3-M_{\chi}^2\chi^*\chi+\lambda_{\chi} (\chi^*\chi)^2+\lambda_1 (\Phi_1^\dagger\Phi_1)^2\nonumber\\
&&\qquad+\lambda_2 (\Phi_2^\dagger\Phi_2)^2+\lambda_3 (\Phi_3^\dagger\Phi_3)^2+\lambda'_4 (\Phi_1^\dagger\Phi_1)(\Phi_2^\dagger\Phi_2)+\lambda_4'' (\Phi_1^\dagger\Phi_2)(\Phi_2^\dagger\Phi_1)\nonumber\\
&&\qquad+\lambda_5' (\Phi_1^\dagger\Phi_1)(\Phi_3^\dagger\Phi_3)+\lambda_5'' (\Phi_1^\dagger\Phi_3)(\Phi_3^\dagger\Phi_1)+\lambda_6' (\Phi_2^\dagger\Phi_2)(\Phi_3^\dagger\Phi_3)+\lambda_6'' (\Phi_2^\dagger\Phi_3)(\Phi_3^\dagger\Phi_2)\nonumber\\
&&\qquad+\lambda_7 (\Phi_1^\dagger\Phi_1)(\chi^*\chi)+\lambda_{8} (\Phi_2^\dagger\Phi_2)(\chi^*\chi)+\lambda_{9} (\Phi_3^\dagger\Phi_3)(\chi^*\chi)+[\lambda_{10} (\Phi_3^\dagger\Phi_1)(\Phi_3^\dagger\Phi_2)\nonumber\\
&&\qquad+\kappa(\Phi_1^\dagger\Phi_2)\chi+h.c.],\label{eqsca}
\end{eqnarray}
where
\begin{eqnarray}
&&\Phi_1=\left(\begin{array}{c}\phi_1^+\\ \frac{1}{\sqrt2}(i A_1+S_1+v_1)\end{array}\right),\Phi_2=\left(\begin{array}{c}\phi_2^+\\ \frac{1}{\sqrt2}(i A_2+S_2+v_2)\end{array}\right),\Phi_3=\left(\begin{array}{c}\phi_3^+\\ \frac{1}{\sqrt2}(i A_3+S_3+v_3)\end{array}\right),\nonumber\\
&&\chi=\frac{1}{\sqrt2}(i A_{\chi}+S_{\chi}+v_\chi),
\end{eqnarray}
$v_i\;(i=1,\;2,\;3),\;v_\chi$ are the vacuum expectation values (VEVs) of $\Phi_i,\;\chi$ respectively, $(v_1^2+v_2^2+v_3^2)^{1/2}= v\approx 246\;{\rm GeV}$ and $v_1,\;v_2< v_3\ll v_\chi$. Based on the scalar potential in Eq.~(\ref{eqsca}), we can obtain four constraint equations from the minimization
\begin{eqnarray}
&&M_{\Phi_1}^2=\lambda_1v_1^2+\frac{1}{2}\Big[(\lambda_4'+\lambda_4'') v_2^2+(\lambda_5'+\lambda_5'') v_3^2+\frac{v_2}{v_1}v_3^2 {\rm Re}(\lambda_{10})+\sqrt2\frac{v_2}{v_1} v_\chi {\rm Re}(\kappa)+\lambda_7v_\chi^2\Big],\nonumber\\
&&M_{\Phi_2}^2=\lambda_2v_2^2+\frac{1}{2}\Big[(\lambda_4'+\lambda_4'') v_1^2+(\lambda_6'+\lambda_6'') v_3^2+\frac{v_1}{v_2}v_3^2 {\rm Re}(\lambda_{10})+\sqrt2\frac{v_1}{v_2} v_\chi {\rm Re}(\kappa)+\lambda_8v_\chi^2\Big],\nonumber\\
&&M_{\Phi_3}^2=\lambda_3v_3^2+{\rm Re}(\lambda_{10})v_1v_2+\frac{1}{2}[(\lambda_5'+\lambda_5'') v_1^2+(\lambda_6'+\lambda_6'') v_2^2+\lambda_9 v_c^2],\nonumber\\
&&M_{\chi}^2=\lambda_\chi v_\chi^2+\frac{1}{2}\Big[\lambda_7 v_1^2+\lambda_8 v_2^2+\lambda_9 v_3^2+\sqrt2\frac{v_1v_2}{v_\chi}  {\rm Re}(\kappa)\Big].\label{eqtad}
\end{eqnarray}

\begin{figure}
\setlength{\unitlength}{1mm}
\centering
\includegraphics[width=1.55in]{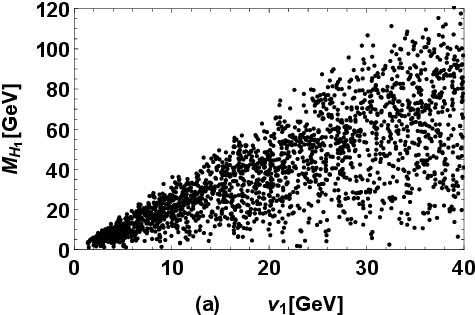}
\vspace{0.2cm}
\includegraphics[width=1.55in]{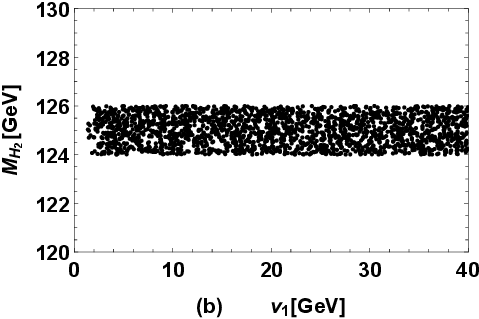}
\vspace{0.2cm}
\includegraphics[width=1.55in]{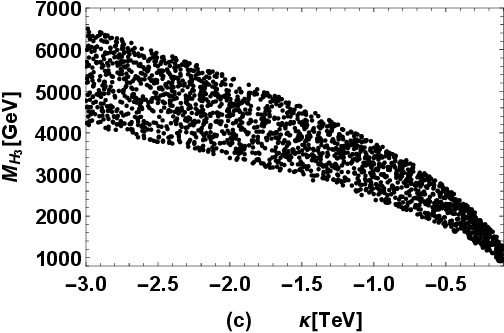}
\vspace{0.2cm}
\includegraphics[width=1.55in]{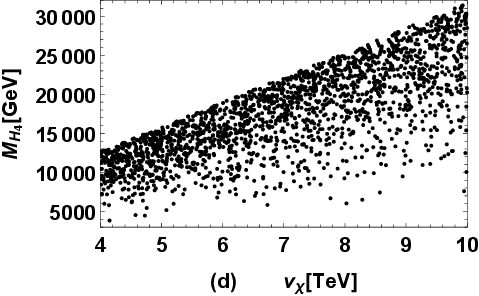}
\vspace{0.2cm}
\includegraphics[width=1.55in]{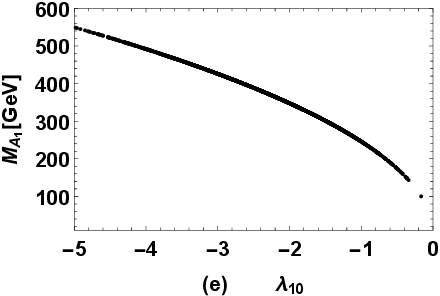}
\vspace{0.2cm}
\includegraphics[width=1.55in]{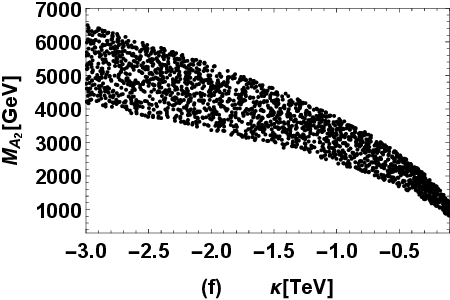}
\vspace{0.2cm}
\includegraphics[width=1.55in]{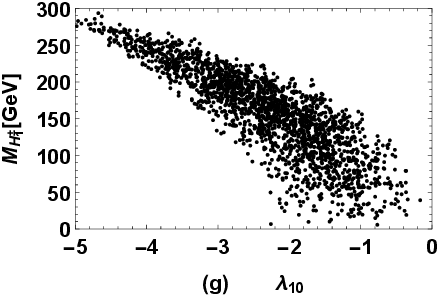}
\vspace{0.2cm}
\includegraphics[width=1.55in]{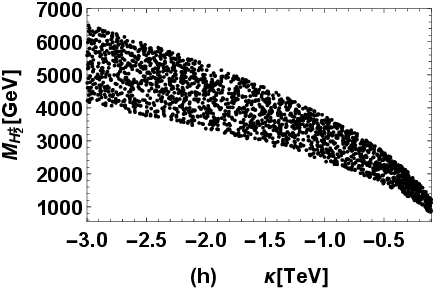}
\vspace{-0.4cm}
\caption[]{Scanning the parameter space in Eq.~(\ref{eq25}), keeping the next-to-lightest CP-even Higgs mass $M_{H_2}$ in the range $124\;{\rm GeV}<M_{H_2}<126\;{\rm GeV}$ and the stability of vacuum, the results of $M_{H_1}$ versus $v_1$ (a), $M_{H_2}$ versus $v_1$ (b), $M_{H_3}$ versus $\kappa$ (c), $M_{H_4}$ versus $v_\chi$ (d), $M_{A_1}$ versus $\lambda_{10}$ (e), $M_{A_2}$ versus $\kappa$ (f), $M_{H_1^\pm}$ versus $\lambda_{10}$ (g) and $M_{H_2^\pm}$ versus $\kappa$ (h) are plotted.}
\label{FigS}
\end{figure}

Firstly, we focus on the Higgs mass spectrum and scalar potential in the FDM which contains four CP-even Higgs, two CP-odd Higgs and two singly charged Higgs. For simplicity, all parameters in Eq.~(\ref{eqsca}) are assumed to be real and $v_1=v_2$, $\lambda_4'=\lambda_4''=\lambda_4/2$, $\lambda_5'=\lambda_5''=\lambda_5/2$, $\lambda_6'=\lambda_6''=\lambda_6/2$. Scanning the following parameter space
\begin{eqnarray}
&&v_1=(0,\;40)\;{\rm GeV},\;\lambda_i=(0,\;5)\;\;{\rm with}\;\;(i=1,...,9,\chi),\;\lambda_{10}=(-5,\;0),\nonumber\\
&&v_\chi=(4,\;10)\;{\rm TeV},\;\kappa=(-3,-0.1)\;{\rm TeV},\label{eq25}
\end{eqnarray}
we plot $M_{H_1}$ versus $v_1$ (a), $M_{H_2}$ versus $v_1$ (b), $M_{H_3}$ versus $\kappa$ (c), $M_{H_4}$ versus $v_\chi$ (d), $M_{A_1}$ versus $\lambda_{10}$ (e), $M_{A_2}$ versus $\kappa$ (f), $M_{H_1^\pm}$ versus $\lambda_{10}$ (g) and $M_{H_2^\pm}$ versus $\kappa$ (h) in Fig.~\ref{FigS}.
In the scanning, we keep the next-to-lightest CP-even Higgs mass $M_{H_2}$ in the range $124\;{\rm GeV}<M_{H_2}<126\;{\rm GeV}$\footnote{The reasons for taking the next-to-lightest CP-even Higgs to be the SM-like Higgs: 1, considering the observed $2.8\sigma$ excess at about $95~$GeV by CMS~\cite{CMS:2023yay} and the observed $2.3\sigma$ local excess at about $98~$GeV by LEP~\cite{LEP:2003ing}, we will explore in our next work whether the additional scalars in the FDM can account for these low-mass excesses; 2, the additional Higgs mass below $125$ GeV by far is not excluded experimentally~\cite{Robens:2022zgk}; 3, if the low-mass Higgs is excluded completely in future, we verify numerically that the next-to-lightest CP-even Higgs mass can reach hundreds GeV by taking the lightest CP-even Higgs mass around 125 GeV and scanning the parameter space in Eq.~(\ref{eq25}).} and the scalar potential at the input $v_1,\;v_2,\;v_3,\;v_\chi$ is smaller than all the other stationary points to guarantee the stability of vacuum.

The right-handed neutrino singlets obtain the Majorana masses after $\chi$ achieving nonzero VEVs. Then the Yukawa coupling in the FDM can be written as
\begin{eqnarray}
&&\mathcal{L}_Y=Y_u^{33}\bar q_3 \tilde \Phi_3 u_{R_3}+Y_d^{33}\bar q_3 \Phi_3 d_{R_3}+Y_u^{32}\bar q_3 \tilde{\Phi}_1 u_{R_2}+Y_u^{23}\bar q_2 \tilde \Phi_1 u_{R_3}+Y_d^{32}\bar q_3 \Phi_2 d_{R_2}\nonumber\\
&&\qquad\; +Y_d^{23}\bar q_2 \Phi_2 d_{R_3}+Y_u^{21}\bar q_2 \tilde{\Phi}_3 u_{R_1}+Y_u^{12}\bar q_1 \tilde \Phi_3 u_{R_2}+Y_d^{21}\bar q_2 \Phi_3 d_{R_1}+ Y_d^{12}\bar q_1 \Phi_3 d_{R_2}\nonumber\\
&&\qquad\; +Y_u^{31}\bar q_3 \tilde{\Phi}_2 u_{R_1}+Y_u^{13}\bar q_1 \tilde \Phi_2 u_{R_3}+Y_d^{31}\bar q_3 \Phi_1 d_{R_1}+Y_d^{13}\bar q_1 \Phi_1 d_{R_3}\nonumber\\
&&\qquad\; +Y_e^{33}\bar l_3 \Phi_3 e_{R_3}+Y_e^{32}\bar l_3 \Phi_2 e_{R_2}+Y_e^{23}\bar l_2 \Phi_2 e_{R_3}+Y_e^{21}\bar l_2 \Phi_3 e_{R_1}+ Y_e^{12}\bar l_1 \Phi_3 e_{R_2}\nonumber\\
&&\qquad\; +Y_e^{31}\bar l_3 \Phi_1 e_{R_1}+Y_e^{13}\bar l_1 \Phi_1 e_{R_3}+Y_R^{11}\bar\nu^c_{R_1}\nu_{R_1}\chi+Y_R^{22}\bar\nu^c_{R_2}\nu_{R_2} \chi^*+Y_D^{21}\bar l_2 \tilde \Phi_3 \nu_{R_1}\nonumber\\
&&\qquad\; +Y_D^{12}\bar l_1 \tilde \Phi_3 \nu_{R_2}+Y_D^{31}\bar l_3 \tilde \Phi_2 \nu_{R_1}+Y_D^{32}\bar l_3 \tilde \Phi_1 \nu_{R_2}+h.c..\label{eq9}
\end{eqnarray}
The mass matrices in Eq.~(\ref{eq2}) can be obtained after $\Phi_1,\;\Phi_2,\;\Phi_3,\;\chi$ receive nonzero VEVs. The gauge boson masses in the model are
\begin{eqnarray}
&&M_\gamma=0,\;M_W=\frac{1}{2}g_2 v,\;M_Z\approx\frac{1}{2}(g_1^2+g_2^2)^{1/2} v,\;M_{Z'}\approx 2|zg_{_F}| v_\chi,\label{eq19}
\end{eqnarray}
where the higher orders of $v/v_\chi$ are neglected for $M_Z$ and $M_{Z'}$.

\begin{figure}
\setlength{\unitlength}{1mm}
\centering
\includegraphics[width=2.7in]{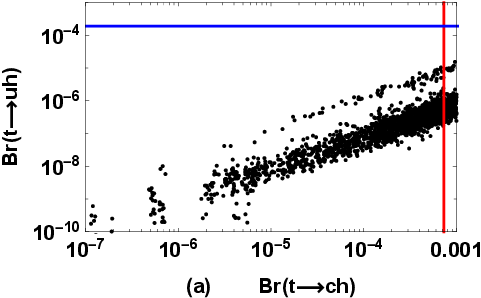}
\vspace{0.2cm}
\includegraphics[width=2.7in]{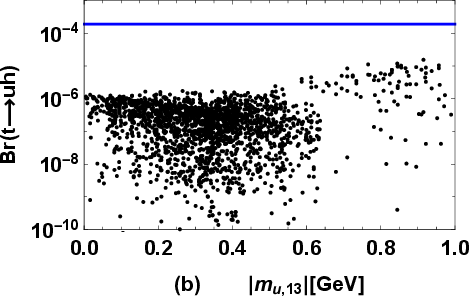}
\vspace{-0.4cm}
\caption[]{Taking the points obtained in Fig.~\ref{Fig1} and Fig.~\ref{FigS} as inputs, the results of ${\rm Br}(t\to uh)-{\rm Br}(t\to ch)$ (a) and ${\rm Br}(t\to uh)-|m_{u,13}|$ are plotted, where the red and blue lines denote the upper bounds on ${\rm Br}(t\to ch)$ and ${\rm Br}(t\to uh)$ from Particle Data Group~\cite{ParticleDataGroup:2022pth} respectively}
\label{Fig2}
\end{figure}
One of the most effective ways to test the FDM is measuring the top quark rare decay processes $t\to ch$ and $t\to uh$, because the tree level coupling between top quark and $H_3$ (which corresponds to the SM Higgs doublet) can make significant contributions to these two processes as shown in Eq.~(\ref{eq9}). The branching ratios can be written as~\cite{YANG2018}
\begin{eqnarray}
&&{\rm Br}(t\rightarrow q_u h)=\frac{|\mathcal{M}_{t q_u h}|^2\sqrt{((m_t+m_h)^2-m_{q_u}^2)((m_t-m_h)^2-m_{q_u}^2)}}{32\pi m_t^3\Gamma^t_{{\rm total}}},
\end{eqnarray}
where $q_u=u,\;c$, the amplitude $\mathcal{M}_{tq_uh}$ can be read directly from the Yukawa couplings in Eq.~(\ref{eq9}), and $\Gamma^t_{{\rm total}}=1.42\;$GeV~\cite{ParticleDataGroup:2022pth} is the total decay width of top quark. Then taking the points obtained in Fig.~\ref{Fig1} and Fig.~\ref{FigS} as inputs, we plot the results of ${\rm Br}(t\to uh)-{\rm Br}(t\to ch)$ in Fig.~\ref{Fig2} (a) and ${\rm Br}(t\to uh)$ versus $|m_{u,13}|$ in Fig.~\ref{Fig2} (b), where the red and blue lines denote the upper bounds on ${\rm Br}(t\to ch)$ and ${\rm Br}(t\to uh)$ from Particle Data Group~\cite{ParticleDataGroup:2022pth} respectively. The picture indicates that the model predicts ${\rm Br}(t\to uh)$ and ${\rm Br}(t\to ch)$ can reach about $10^{-5}$ and $10^{-4}$ respectively, which have great opportunities to be observed in the next generation of experiments. In addition, large ${\rm Br}(t\to uh)$ prefers large $|m_{u,13}|$ ($|m_{u,13}|\gtrsim0.65\;{\rm GeV}$) as shown in Fig.~\ref{Fig2} (b) while large $|m_{u,13}|$ is hard to fit the measured CKM matrix elements.

Due to the introducing of flavor dependent gauge symmetry $U(1)_F$, $Z$ and new $Z'$ bosons in the FDM can make contributions at the tree level to the lepton flavor violation (LFV) processes $\mu\to3e,\;\tau\to3e,\;\mu\to3\mu$, which suffer strict experimental constraints ${\rm Br}(\mu\to3e)<1\times 10^{-12}$~\cite{SINDRUM:1987nra}, ${\rm Br}(\tau\to3e)<2.7\times 10^{-8}$~\cite{Hayasaka:2010np}, ${\rm Br}(\tau\to3\mu)<2.1\times 10^{-8}$~\cite{Hayasaka:2010np}. In addition, Higgs bosons can also make contributions to these LFV processes at the tree level, we take $|m_{e,13}|=0.01\;{\rm GeV}$, $\theta_{e,12}=\theta_{e,13}=\theta_{e,23}=0$ and the $U(1)_F$ charge $z=1$ for simplicity. The corresponding amplitude can be written as~\cite{Hisano:1995cp}
\begin{eqnarray}
&&\mathcal{M}(e_j\rightarrow e_i e_i\bar e_i)=C_1^L\bar u_{e_i}(p_2)\gamma_\mu P_L u_{e_j}(p_1) u_{e_i}(p_3)\gamma^\mu P_L \nu_{e_i}(p_4)\nonumber\\
&&\qquad\quad+C_1^R\bar u_{e_i}(p_2)\gamma_\mu P_R u_{e_j}(p_1) u_{e_i}(p_3)\gamma^\mu P_R \nu_{e_i}(p_4)\nonumber\\
&&\qquad\quad+[C_2^L\bar u_{e_i}(p_2)\gamma_\mu P_L u_{e_j}(p_1) u_{e_i}(p_3)\gamma^\mu P_R \nu_{e_i}(p_4)\nonumber\\
&&\qquad\quad+C_2^R\bar u_{e_i}(p_2)\gamma_\mu P_R u_{e_j}(p_1) u_{e_i}(p_3)\gamma^\mu P_L \nu_{e_i}(p_4)-(p_2\leftrightarrow p_3)]\nonumber\\
&&\qquad\quad+[C_3^L\bar u_{e_i}(p_2) P_L u_{e_j}(p_1) u_{e_i}(p_3) P_L \nu_{e_i}(p_4)\nonumber\\
&&\qquad\quad+C_3^R\bar u_{e_i}(p_2) P_R u_{e_j}(p_1) u_{e_i}(p_3) P_R \nu_{e_i}(p_4)-(p_2\leftrightarrow p_3)],
\end{eqnarray}
where $i=1,\;2$ for $j=3$, $i=1$ for $j=2$, $u_{e_i}$ denotes the spinor of lepton, $\nu_{e_i}$ denotes the spinor of antilepton, $P_L=(1-\gamma_5)/2$, $P_R=(1+\gamma_5)/2$, and $p_k$ denotes the momentum of charged lepton with $k=1,2,3,4$. The coefficients $C_{1,2,3}^{L,R}$ from the contributions of Higgs bosons and $Z,\;Z'$ gauge bosons, can be obtained through the Yukawa couplings in Eq.~(\ref{eq9}) and the definition of covariant derivative corresponding to $SU(2)_L\otimes U(1)_Y\otimes U(1)_F$
\begin{eqnarray}
&&D_\mu=\partial_\mu+i g_2 T_a A_{\mu}^a+i g_1 Y B_\mu+i g_{_F} F B'_\mu+i g_{_{YF}} Y B'_\mu,\;(a=1,2,3)\label{eqD}
\end{eqnarray}
respectively. In Eq.~(\ref{eqD}), $g_{_{YF}}$ is the gauge coupling constant arises from the gauge kinetic mixing effect, $(g_2,\;g_1,\; g_{_F})$, $(T_a,\;Y,\;F)$, $(A_{a\mu},\;B_\mu,\; B'_\mu)$ denote the gauge coupling constants, generators and gauge bosons of groups $(SU(2)_L,\;U(1)_Y,\;U(1)_F)$ respectively. Then we can calculate the decay rate~\cite{Hisano:1995cp}
\begin{eqnarray}
&&\Gamma(e_j\rightarrow e_i e_i\bar e_i)=\frac{m_{e_j}^5}{1536\pi^3}\Big[\frac{1}{2}(|C_1^L|^2+|C_1^R|^2)+|C_2^L|^2+|C_2^R|^2+\frac{1}{8}(|C_3^L|^2+|C_3^R|^2)\Big].
\end{eqnarray}
The total decay widthes of $\mu,\;\tau$ are taken as $\Gamma^\mu_{{\rm total}}=2.996\times 10^{-19}\;$GeV, $\Gamma^\tau_{{\rm total}}=2.265\times 10^{-12}\;$GeV~\cite{ParticleDataGroup:2022pth}.

\begin{figure}
\setlength{\unitlength}{1mm}
\centering
\includegraphics[width=2.1in]{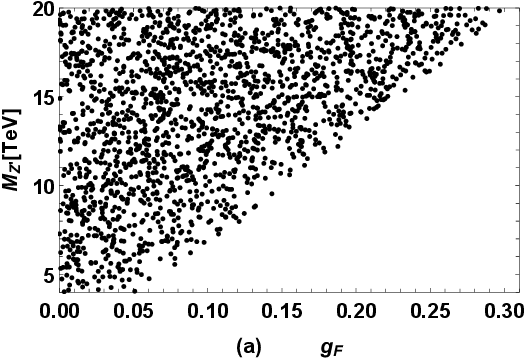}
\vspace{0.2cm}
\includegraphics[width=2.1in]{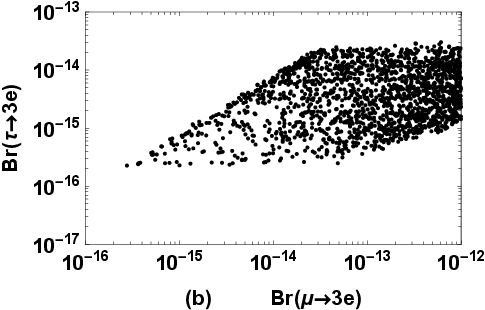}
\vspace{0.2cm}
\includegraphics[width=2.1in]{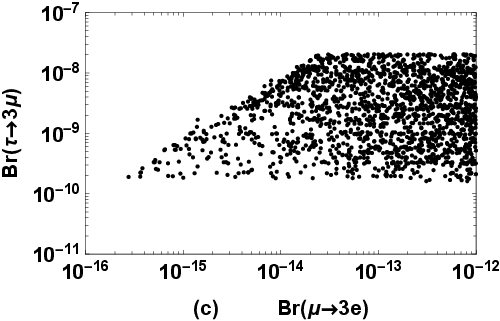}
\vspace{-0.4cm}
\caption[]{Taking the points obtained in Fig.~\ref{Fig1}, Fig.~\ref{FigS} as inputs, and keeping ${\rm Br}(\mu\to3e)<1\times 10^{-12}$, ${\rm Br}(\tau\to3e)<2.7\times 10^{-8}$, ${\rm Br}(\tau\to3\mu)<2.1\times 10^{-8}$ in the computation, the allowed ranges of $M_{Z'}-g_F$ (a) and the results of ${\rm Br}(\tau\to 3e)-{\rm Br}(\mu\to 3e)$ (b), ${\rm Br}(\tau\to 3\mu)-{\rm Br}(\mu\to 3e)$ (c) are plotted.}
\label{Fig3}
\end{figure}

Taking the points obtained in Fig.~\ref{Fig1}, Fig.~\ref{FigS} as inputs, and keeping ${\rm Br}(\mu\to3e)<1\times 10^{-12}$, ${\rm Br}(\tau\to3e)<2.7\times 10^{-8}$, ${\rm Br}(\tau\to3\mu)<2.1\times 10^{-8}$ in the computation, we plot the allowed ranges of $M_{Z'}-g_F$ in Fig.~\ref{Fig3} (a), the results of ${\rm Br}(\tau\to 3e)-{\rm Br}(\mu\to 3e)$ in Fig.~\ref{Fig3} (b), the results of ${\rm Br}(\tau\to 3\mu)-{\rm Br}(\mu\to 3e)$ in Fig.~\ref{Fig3} (c).
It is obvious from Fig.~\ref{Fig3} (a) that the experimental upper bounds on the branching ratios of these LFV processes give the constraints $M_{Z'}/g_F\gtrsim70\;{\rm TeV}$ in our chosen parameter space. And Fig.~\ref{Fig3} (b), Fig.~\ref{Fig3} (c) show that ${\rm Br}(\mu\to3e)\gtrsim 10^{-16}$, ${\rm Br}(\tau\to3e)\gtrsim 10^{-16}$, ${\rm Br}(\tau\to3\mu)\gtrsim 10^{-10}$ in the FDM, which have great opportunities to be observed in the future~\cite{Blondel:2013ia,Hayasaka:2013dsa}.

Focusing on the hierarchical structure of fermionic masses puzzle and fermionic flavor mixings puzzle, we propose to relate these two puzzles by the see-saw mechanism, i.e. only the third generation of quarks and charged leptons (the SM neutrinos obtain tiny Majorana masses through the so-called Type-I see-saw mechanism by introducing the right-handed neutrinos) achieve the masses at the tree level, and the first two generations achieve masses through the mixings with the third generation. Taking the pole quark masses as inputs, we found that the mechanism proposed in this work can well fit the measured CKM matrix elements. Finally, a flavor-dependent model (FDM) are proposed in this letter to realize the new mechanism mentioned above. The model extends the SM by an extra $U(1)_F$ local gauge group, two scalar doublets and two scalar singlets, where the new $U(1)_F$ charges are related to the particles' flavor. And we point out that observing the top quark rare decay processes $t\to ch$, $t\to uh$ and the lepton flavor violation processes $\mu\to3e,\;\tau\to3e,\;\mu\to3\mu$ is an effective way to test the FDM.

\begin{acknowledgments}

The work has been supported by the National Natural Science Foundation of China (NNSFC) with Grants No. 12075074, No. 12235008, Hebei Natural Science Foundation with Grant No. A2022201017, No. A2023201041, Natural Science Foundation of Guangxi Autonomous Region with Grant No. 2022GXNSFDA035068, the youth top-notch talent support program of the Hebei Province.

\end{acknowledgments}

\end{document}